# THERMODYNAMIC PROCESSES GENERATED BY A CLASS OF COMPLETELY POSITIVE QUANTUM OPERATIONS

SUMIYOSHI ABE* and YUKI AOYAGHI

*Department of Physical Engineering, Mie University, Mie 514-8507, Japan*

* suabe@sf6.so-net.ne.jp

An attempt toward the operational formulation of quantum thermodynamics is made by employing the recently proposed operations forming positive operator-valued measures for generating thermodynamic processes. The quantity of heat as well as the von Neumann entropy monotonically increases under the operations. The fixed point analysis shows that repeated applications of these operations to a given system transform from its pure ground state at zero temperature to the completely random state in the high temperature limit with intermediate states being generically out of equilibrium. It is shown that the Clausius inequality can be violated along the processes, in general. A bipartite spin-1/2 system is analyzed as an explicit example.

___________________________________________

* Corresponding author.

**1. Introduction**

Recently, there has been a great interest in understanding possible roles of quantum mechanics in thermodynamics.[1] This stream seems to physically originate in nanoscience and quantum information/computation. In the former, it is necessary to take into account quantum fluctuations due to smallness of the system size, while in reality the environmental effects are inevitable in the latter.

It is generally assumed in statistical mechanics that a system surrounded by its environment in an arbitrary initial state approaches equilibrium in an irreversible way, although the underlying microscopic dynamics is reversible. There is a traditional opinion that this nonunitary nature is an indication of possible fundamental relevance of problems of quantum measurement to quantum thermodynamics.[2] That is, the effects of the environment may be interpreted as those of measurement.

In studies of quantum open systems, it is traditional to start the discussion with the total isolated system governed by unitary dynamics, divide it into the objective system and the environment that are weakly interacting each other and then consider nonunitary subdynamics of the objective system.[3] It is, however, known[4,5] that the reduced density matrix of the objective system, obtained from the total density matrix through the partial trace over the environmental degrees of freedom, violates positive semi-definiteness, in general, leading to the fundamental difficulty of the probabilistic interpretation. On the other hand, in classical thermodynamics, the role of the heat bath as an environment is not explicitly dynamical. The heat bath is commonly treated in an operational manner such as exchange of heat with the objective system. This observation makes it natural to formulate an operational approach also to quantum systems.

Here, we address ourselves to developing the first step toward the operational formulation of quantum thermodynamics. Our idea is to replace the environmental effects by nonunitary positive quantum operations in the space of the objective system. In particular, we employ the statistical quantum operation recently proposed in Refs. 6 and 7. Both the von Neumann entropy and the quantity of heat are proven to monotonically increase under this operation. Performing the fixed point analysis, we find that the repeated applications of the operation transform the pure ground state of the system at zero temperature to the completely random state at infinite temperature with intermediate states being generically out of equilibrium. Identifying the series of changes of a quantum state by the repeated operations with a thermodynamic process, we show that the Clausius inequality can be violated along such a process, in general. A bipartite spin-1/2 system is analyzed in detail as a simple and illustrative example.

This paper is organized as follows. In the next section, we recapitulate the statistical quantum operations proposed in Refs. 6 and 7. In Sec. 3, we perform the fixed point analysis to show how the repeated applications of the operation transform the ground state at zero temperature to the completely random state at infinite temperature. In Sec. 4, we show how the von Neumann entropy and the quantity of heat monotonically increase by the operations. Then, in Sec. 5, a thermodynamics process generated by a small operation is discussed for a bipartite spin-1/2 system as an example. There, the Clausius inequality is found to be violated along the process, in general. Section 6 is devoted to concluding remarks.

## 2. Statistical Quantum Operation

Let us employ a quantum system with a Hamiltonian, $H$, in $d$ dimensions, where $d$ can be very large. The collection of the normalized energy eigenstates, $\{|u_n\rangle\}_{n=0,1,...,d-1}$ satisfying $H|u_n\rangle = \varepsilon_n |u_n\rangle$ with the energy eigenvalue $\varepsilon_n$, is assumed to form a complete orthonormal system:

$$I = \sum_{n=0}^{d-1} |u_n\rangle\langle u_n|, \qquad (1)$$

where $I$ is the $d \times d$ identity matrix.

A quantum operation, $\Phi$, is a map from a density matrix of the system, $\rho$, to another: $\rho \to \Phi(\rho)$. On $\Phi$, we impose the following three conditions: (i) linear, (ii) completely positive and (iii) trace-preserving. (i) is a characteristic of quantum dynamics. (ii) combined with (iii) can assure the probabilistic interpretation. The most general form of $\Phi$ satisfying these three conditions is known to have the following form [8]:

$$\rho \rightarrow \Phi(\rho) = \sum_{n=0}^{d-1} V_n \rho V_n^\dagger, \qquad (2)$$

where the operators $V_n$'s obey

$$\sum_{n=0}^{d-1} V_n^\dagger V_n = I, \qquad (3)$$

which assures the trace-preserving condition: $\text{Tr}\,\Phi(\rho) = \text{Tr}\,\rho\,(=1)$. $\{V_n^\dagger V_n\}_{n=0,1,\ldots,d-1}$ is then called a positive operator-valued measure (POVM).[8,9]

A number of discussions have been made about operations of the type in Eq. (2) in the literature. However, most of them are formal. They are mainly concerned with measurement problems, and, to our knowledge, no studies have been developed in the thermodynamic context.

Here, we base our discussion on the statistical quantum operations presented in Refs. 6 and 7. A statistical state involves the mixture of all the relevant states. Therefore, the transitions between $|u_0\rangle$ and the other eigenstates should be introduced. The basic idea is to pick up two terms in Eq. (1), $|u_0\rangle\langle u_0|$ and $|u_n\rangle\langle u_n|$, and replace them with the transition operators, $|u_0\rangle\langle u_n|$ and $|u_n\rangle\langle u_0|$. Thus, we come to consider the following operators:

$$V_n = a_n \left( I - |u_0\rangle\langle u_0| - |u_n\rangle\langle u_n| + |u_0\rangle\langle u_n| + |u_n\rangle\langle u_0| \right), \tag{4}$$

where $n = 0, 1, ..., d-1$, and $a_n$ is a complex *c*-number that is referred to here as the "transition amplitude". Henceforth, $\Phi$ implies Eq. (2) with Eq. (4).

A straightforward calculation shows that the set of the operators in Eq. (4) satisfies Eq. (3), if the transition amplitude is normalized as

$$\sum_{n=0}^{d-1} |a_n|^2 = 1. \tag{5}$$

Let us apply the above operation to the pure ground state, $|u_0\rangle\langle u_0|$, which is the state of the system at zero temperature. After some calculations, one finds

$$\Phi(|u_0\rangle\langle u_0|) = \sum_{n=0}^{d-1} p_n |u_n\rangle\langle u_n|, \tag{6}$$

where

$$p_n = |a_n|^2. \tag{7}$$

This is a desirable property, since both perfect decoherence (i.e., the absence of the off-diagonal elements in the energy eigenbasis) and statistical mixture are simultaneously realized. In the particular case when $p_n$ is chosen to be the canonical form, $p_n = \exp(-\beta \varepsilon_n)/Z(\beta)$ with the partition function, $Z(\beta) = \text{Tr}\exp(-\beta H) = \sum_{n=0}^{d-1} \exp(-\beta \varepsilon_n)$, and the inverse temperature, $\beta = 1/(k_B T)$ ($k_B$ being the Boltzmann constant), then $\Phi$ becomes an operation that transforms from the zero-temperature ground state, $|u_0\rangle\langle u_0|$, to an equilibrium state at finite temperature, $\rho_{eq} = \exp(-\beta H)/Z(\beta)$. However, clearly, it is not necessary to limit oneself to such a special choice, in general. These are the characteristics of the present operation $\Phi$, which do not seem to have been noticed in the literature.

Closing this section, we wish to mention that the form of the operators in Eq. (4) implicitly indicates the existence of entanglement between the objective and (hidden) environmental systems. This can be seen as follows. Recall Kraus' original discussion.[10] Change of the total system can be described by a certain unitary transformation: $\rho_{tot} \to U \rho_{tot} U^\dagger$. Kraus' choice for the initial total density matrix is $\rho_{tot} = \rho \otimes |\psi\rangle_E {}_E\langle\psi|$, where $\rho$ and $|\psi\rangle_E {}_E\langle\psi|$ are the density matrices of the objective and environmental systems, respectively. Note that the total density matrix is factorized and the environment is in a pure state. No entanglement exists between the two. Taking the partial trace over the environmental degrees of freedom, one obtains the transformation of the reduced density matrix of the objective system: $\text{Tr}_E(U \rho_{tot} U^\dagger) = \sum_n W_n \rho W_n^\dagger$, where $W_n = {}_E\langle v_n|U|\psi\rangle_E$ with $\{|v_n\rangle_E\}_n$ being a certain basis of the Hilbert space of the environment. Now, it is of crucial importance to note that the operators in Eq. (4) does not have this form of $W_n$. In particular, the initial structure, $\rho_{\text{tot}} = \rho \otimes |\psi\rangle_E {}_E\langle\psi|$, is not realized. In other words, the form of the operators in Eq. (4) implicitly indicates that there exists initial entanglement between the objective and environmental systems. This may drastically change the classical concepts in thermodynamics.

## 3. Fixed Point Analysis

The present quantum operations further possess remarkable properties. In this section, we consider repeated applications of the operations.

Let us apply $\Phi$ to Eq. (6). Using the formula

$$\Phi(|u_k\rangle\langle u_k|) = |u_k\rangle\langle u_k| + p_k(|u_0\rangle\langle u_0| - |u_k\rangle\langle u_k|) \quad (k=1,2,...,d-1) \quad (8)$$

as well as Eq. (6), we have

$$\Phi^2(|u_0\rangle\langle u_0|) = \sum_{n=0}^{d-1} p_n^{(2)} |u_n\rangle\langle u_n|, \quad (9)$$

where $p_0^{(2)} = \sum_{n=0}^{d-1} p_n^2$ and $p_k^{(2)} = p_k p_0 + (1-p_k)p_k$ ($k=1,2,...,d-1$). Therefore, the perfect decoherence is preserved through repeated applications of $\Phi$.

Since the system resides in a large but finite dimensional space, $\Phi$ must have its fixed points. Let us write

$$\Phi^N(|u_0\rangle\langle u_0|) = \sum_{n=0}^{d-1} p_n^{(N)} |u_n\rangle\langle u_n|. \tag{10}$$

Applying $\Phi$ to both sides of this equation, we have the following relations:

$$p_0^{(N+1)} = \sum_{n=0}^{d-1} p_n p_n^{(N)}, \tag{11}$$

$$p_k^{(N+1)} = p_k p_0^{(N)} + (1-p_k) p_k^{(N)} \qquad (k=1,2,...,d-1), \tag{12}$$

which, in the limit $N \to \infty$, become

$$p_0^{(\infty)} = \sum_{n=0}^{d-1} p_n p_n^{(\infty)}, \tag{13}$$

$$p_k^{(\infty)} = p_k p_0^{(\infty)} + (1-p_k) p_k^{(\infty)} \qquad (k=1,2,...,d-1), \tag{14}$$

respectively. The coupled Eqs. (13) and (14) have two solutions. One is $p_0 = 1$ and $p_k = 0$ ($k=1,2,...,d-1$), which is trivial since $\Phi$ in this case is the identity operation. The other solution is nontrivial:

$$p_n^{(\infty)} = \frac{1}{d} \qquad (n=0,1,...,d-1). \tag{15}$$

This solution gives rise to

$$\Phi^\infty\left(|u_0\rangle\langle u_0|\right) = \frac{1}{d}I. \tag{16}$$

Eq. (16) implies that the repeated applications of $\Phi$ transforms from the pure ground state at zero temperature to the completely random state realized at infinite temperature. This fact indicates that $\Phi$ is a heat-up operation. We shall show in the next section that this is in fact the case.

4. Entropy and Heat

Here, first let us reexamine the result in Sec. 3 in a somewhat mathematical manner. What is crucial is the fact that $\Phi$ is *unital*.[11,12] implying that the identity operator, $I$, is a fixed point of $\Phi$: $\Phi(I) = I$. In other words, the operators in Eq. (4) satisfies not only Eq. (3) but also

$$\sum_{n=0}^{d-1} V_n V_n^\dagger = I. \tag{17}$$

This equation holds since the operator, $V_n$, is normal, i.e., $[V_n, V_n^\dagger] = 0$. A unital operation has an important property. Let $f$ be operator concave. That is, for any Hermitian operators, $A$ and $B$, it satisfies $f(\lambda A + (1-\lambda)B) \geq \lambda f(A) + (1-\lambda) f(B)$, where $\lambda \in [0,1]$. (The operator inequality, $A \geq B$, means that all the eigenvalues of $A - B$ are nonnegative.) Then, unital $\Phi$ satisfies[11,12]

$$f(\Phi(A)) \geq \Phi(f(A)). \tag{18}$$

Choose $A$ to be an arbitrary density matrix, $\rho$, and set

$$f(\rho) = -k_B \rho \ln \rho, \tag{19}$$

which is strictly operator concave. Then, taking the trace of Eq. (18), we obtain

$$S[\Phi(\rho)] \geq S[\rho], \tag{20}$$

where $S$ is the von Neumann entropy defined by

$$S[\rho] = -k_B \text{Tr}(\rho \ln \rho). \tag{21}$$

Therefore, the entropy monotonically increases under the application of $\Phi$, if $\rho$ is not a fixed point of $\Phi$. Thus,

$$S\left[\Phi^N(\rho)\right] \xrightarrow{N \to \infty} S_{\max} = k_B \ln d, \tag{22}$$

where $S_{\max}$ is the maximum value of the von Neumann entropy of the completely random state in Eq. (16). Note that, in the above discussion, $\rho$ is an arbitrary density matrix and is not necessarily equal to $|u_0\rangle\langle u_0|$, unlike in the preceding section.

Next, let us consider the quantity of heat. The change of the internal energy,

$U = \text{Tr}(\rho H)$, along a certain thermodynamic process is given by $\delta U = \text{Tr}(H \delta \rho) + \text{Tr}(\rho \delta H)$. Identifying the first and second terms on the right-hand side with the changes of the quantity of heat, $\delta'Q$, and work, $-\delta'W$, we recognize the first law of thermodynamics:

$$\delta'Q = \delta U + \delta'W. \tag{23}$$

In the above identifications, it is essential to note that $-\delta H$ is the work operator, and its expectation value is $\delta'W$. Thus, the quantity, $\text{Tr}(H \delta \rho)$, is necessarily identified with the change of the quantity of heat.

Here, we are interested in a thermodynamic process generated by $\Phi$. The change of the state is thus given by

$$\delta^{(N)} \rho = \Phi^{N+1}(\rho) - \Phi^N(\rho) \qquad (N = 0, 1, 2, ...), \tag{24}$$

provided that $\Phi^0(\rho) \equiv \rho$ (see Fig. 1). Accordingly, we have the following expression for the change of the quantity of heat:

$$\delta'^{(N)} Q = \text{Tr}\left\{ H \left[ \Phi^{N+1}(\rho) - \Phi^N(\rho) \right] \right\}. \tag{25}$$

For $\rho = |u_0\rangle\langle u_0|$ as in Sec. 2, this quantity is rewritten as

$$\delta'^{(N)} Q = \sum_{n=0}^{d-1} \varepsilon_n \left[ p_n^{(N+1)} - p_n^{(N)} \right], \qquad (26)$$

where the notation in Eq. (10) is used. From Eqs. (11) and (12), this quantity is further rewritten as follows:

$$\delta'^{(N)} Q = \sum_{k=1}^{d-1} (\varepsilon_k - \varepsilon_0) p_k \left[ p_0^{(N)} - p_k^{(N)} \right]. \qquad (27)$$

Therefore, if

$$p_0^{(N)} \geq p_k^{(N)} \qquad (k = 1, 2, ..., d-1), \qquad (28)$$

then

$$\delta'^{(N)} Q \geq 0 \qquad (29)$$

holds. Note that Eq. (28) is natural, since it means that, along the thermodynamic process, the probability of finding the system in the ground state is equal to or larger than that in an excited state. In the limit $N \to \infty$, $\delta'^{(N)} Q$ tends to vanish, because of the convergence to the fixed point described by Eq. (15).

Eq. (29) establishes the fact that $\Phi$ is a heat-up operation as long as the natural condition in Eq. (28) is satisfied, as promised in the preceding section.

## 5. Example of Bipartite Spin-1/2 System

Although a bipartite spin-1/2 system is very small from the thermodynamic viewpoint, it may cast further light on the physical nature of the present approach to quantum thermodynamics.

The Hamiltonian we employ here is of the Heisenberg type:

$$H = -J \sigma_A \cdot \sigma_B \quad (30)$$

with an antiferromagnetic coupling constant

$$J < 0, \quad (31)$$

where $\sigma_\alpha = (\sigma_{\alpha,x}, \sigma_{\alpha,y}, \sigma_{\alpha,z})$ ($\alpha = A, B$) are the Pauli matrices of the two spins, $A$ and $B$. The eigenstates of this Hamiltonian are given by

$$|u_0\rangle = \frac{1}{\sqrt{2}} \left( |\uparrow\rangle_A |\downarrow\rangle_B - |\downarrow\rangle_A |\uparrow\rangle_B \right), \quad (32)$$

$$|u_1\rangle = \frac{1}{\sqrt{2}} \left( |\uparrow\rangle_A |\downarrow\rangle_B + |\downarrow\rangle_A |\uparrow\rangle_B \right), \quad (33)$$

$$|u_2\rangle = \frac{1}{\sqrt{2}} \left( |\uparrow\rangle_A |\uparrow\rangle_B - |\downarrow\rangle_A |\downarrow\rangle_B \right), \quad (34)$$

$$|u_3\rangle = \frac{1}{\sqrt{2}} \left( |\uparrow\rangle_A |\uparrow\rangle_B + |\downarrow\rangle_A |\downarrow\rangle_B \right), \quad (35)$$

where $|\uparrow\rangle_\alpha$ and $|\downarrow\rangle_\alpha$ ($\alpha = A, B$) are the eigenstates of $\sigma_{\alpha,z}$ corresponding to the eigenvalues, $+1$ and $-1$, respectively. These states are the maximally entangled states termed the Bell states. They form a complete orthonormal system in the 4-dimensional space. The energy eigenvalues are:

$$\varepsilon_0 = 3J, \qquad \varepsilon_1 = \varepsilon_2 = \varepsilon_3 = -J. \tag{36}$$

The excited state is three-fold degenerate.

A quantum operation, $\Phi$, we consider here is characterized by

$$p_0 = 1 - \Delta, \qquad p_k = \frac{\Delta}{3} \quad (k = 1, 2, 3). \tag{37}$$

This $\Phi$ is a heat-up operation if Eq. (28) is satisfied. In the present case, it reads $0 < \Delta \leq 3/4$. In what follows, we assume that $\Delta$ is a positive small constant.

So, let us consider a thermodynamic process generated by the operation, $\Phi$, with Eq. (37). We start with the canonical density matrix of an equilibrium state

$$\rho_{eq} = \frac{1}{Z(\beta)} \exp(\beta J \sigma_A \cdot \sigma_B)$$

$$= \sum_{n=0}^{3} f_n |u_n\rangle\langle u_n|, \tag{38}$$

where $Z(\beta)$ is the partition function

$$Z(\beta) = e^{-3\beta J} + 3e^{\beta J} \qquad (39)$$

and $f_n$'s are the probabilities

$$f_0 = \frac{e^{-3\beta J}}{Z(\beta)}, \qquad f_1 = f_2 = f_3 = \frac{e^{\beta J}}{Z(\beta)}, \qquad (40)$$

which satisfy the normalization condition: $f_0 + 3f_k = 1$ ($k=1, 2, 3$). We repeatedly apply $\Phi$ to this "initial state", i.e., Fig. 1 with $\rho = \rho_{eq}$. The change of the quantity of heat along this process is given by

$$\delta'^{(N)}Q = \text{Tr}\left\{ H\left[ \Phi^{N+1}(\rho_{eq}) - \Phi^N(\rho_{eq}) \right] \right\}. \qquad (41)$$

On the other hand, there seem to be two possibilities for defining the change of the entropy:

$$\delta_{\text{I}}^{(N)}S = S\left[\Phi^{N+1}(\rho_{eq})\right] - S\left[\Phi^N(\rho_{eq})\right], \qquad (42)$$

$$\delta_{\text{II}}^{(N)}S = -k_B \text{Tr}\left\{ \left[ \Phi^{N+1}(\rho_{eq}) - \Phi^N(\rho_{eq}) \right] \ln \Phi^N(\rho_{eq}) \right\}. \qquad (43)$$

The second definition comes from the direct variation of the von Neumann entropy: $\delta S[\rho] = -k_B \text{Tr}(\delta \rho \ln \rho)$ with the identification in Eq. (24). In what follows, we examine both of these definitions. The reason why we consider both of them is as follows. They coincide with each other up to the first order of $\delta\Phi^N(\rho) = \Phi^{N+1}(\rho) - \Phi^N(\rho)$. However, the higher-order contributions, which make Eqs. (42) and (43) different from each other, turn out to be important in the subsequent discussion.

We are going to consider the thermodynamic process generated by $\Phi$ around the equilibrium state. We have

$$\Phi(\rho_{eq}) = \rho_{eq} + \frac{\Delta}{3}(f_0 - f_1)\left(I - 4|u_0\rangle\langle u_0|\right), \tag{44}$$

$$\Phi^2(\rho_{eq}) = \rho_{eq} + \left(\frac{2\Delta}{3} - \frac{4\Delta^2}{9}\right)(f_0 - f_1)\left(I - 4|u_0\rangle\langle u_0|\right), \tag{45}$$

which are not equilibrium states any more. Using these quantities, we can calculate Eq. (41) as follows:

$$\delta'^{(0)}Q = -4\Delta J(f_0 - f_1), \tag{46}$$

$$\delta'^{(1)}Q = -\left(4\Delta - \frac{16\Delta^2}{3}\right)J(f_0 - f_1). \tag{47}$$

On the other hand, Eqs. (42) and (43) are calculated as follows:

$$\delta_I^{(0)}S = -4\Delta k_B \beta J(f_0 - f_1) - \frac{\Delta^2}{6}k_B \frac{(f_0 - f_1)^2}{f_0 f_1} + O(\Delta^3), \tag{48}$$

$$\delta_{\mathrm{I}}^{(1)}S = -4\Delta k_B \beta J(f_0 - f_1)$$
$$+\frac{\Delta^2}{6}k_B(f_0 - f_1)\left[32\beta J - 3\left(\frac{1}{f_1} - \frac{1}{f_0}\right)\right] + O(\Delta^3), \tag{49}$$

and

$$\delta_{\mathrm{II}}^{(0)}S = -4\Delta k_B \beta J(f_0 - f_1), \tag{50}$$

$$\delta_{\mathrm{II}}^{(1)}S = -4\Delta k_B \beta J(f_0 - f_1)$$
$$+\frac{\Delta^2}{3}k_B(f_0 - f_1)\left[16\beta J - \left(\frac{1}{f_1} - \frac{1}{f_0}\right)\right] + O(\Delta^3), \tag{51}$$

respectively.

From Eqs. (46) and (50), we find that the Clausius equality

$$\delta_{\mathrm{II}}^{(0)}S = \frac{\delta'^{(0)}Q}{T}, \tag{52}$$

holds, describing the reversibility of the infinitesimal process around the equilibrium state, if $\delta_{\mathrm{II}}^{(0)}S$ is employed. Note, however, that such an equality does not hold if $\delta_{\mathrm{I}}^{(0)}S$ is used. In fact, we have

$$\delta_{\mathrm{I}}^{(0)}S - \frac{\delta'^{(0)}Q}{T} = -\frac{\Delta^2}{6}k_B\frac{(f_0 - f_1)^2}{f_0 f_1} + O(\Delta^3), \tag{53}$$

which is negative. Likewise, comparing $\delta_{\mathrm{II}}^{(1)}S$ with $\delta'^{(1)}Q$, we have

$$\delta_{\text{II}}^{(1)} S - \frac{\delta'^{(1)} Q}{T} = -\frac{\Delta^2}{3} k_B \frac{(f_0 - f_1)^2}{f_0 f_1} + O(\Delta^3) , \qquad (54)$$

which is also negative.

Eqs. (53) and (54) indicate that the Clausius inequality, $\delta S \geq \delta' Q / T$, is violated in the thermodynamics process considered here, in general. The violation is of the higher order, $O(\Delta^2)$, which may imply that, around the equilibrium state, the system still prefers to experience a reversible process.

It is noted that the negativities in both Eqs. (53) and (54) are proportional to $(f_0 - f_1)^2$. That is, the larger the energy gap is, the stronger the violation of the Clausius inequality is. So, the quantum discreteness is essential for the negativities. A more important point is concerned with the discussion made in the closing part of Sec. 2. As mentioned there, the present quantum operations cannot be expressed in the form of the original Kraus representation[10], in which the quantum states of the objective and environmental systems are initially factorized. Therefore, quantum entanglement between the systems must also play a crucial role for the violation of the Clausius inequality. The present result as well as these observations may remind one of recent discussions[13-15] about quantum-mechanical violation of the second law of thermodynamics. Clearly, further investigations are needed for understanding these issues.

## 6. Conclusion

We have studied the recently-proposed positive quantum operations[6,7] from the thermodynamic viewpoints. We have performed the fixed point analysis and have found that the operations transform from the pure ground state at zero temperature to the completely random state at infinite temperature as the nontrivial fixed point of the operations. We have shown that both the von Neumann entropy and the quantity of heat monotonically increase by the operations. Thus, the operations have been proven to heat up systems. As an analytically tractable example, we have considered the bipartite spin-1/2 system. We have found that the Clausius inequality may be violated along the thermodynamic process generated by the present quantum operations, in general.

**Acknowledgement**

The work of S. A. was supported in part by a Grant-in-Aid for Scientific Research from the Japan Society for the Promotion of Science.

**Figure Caption**

Fig. 1. A schematic description of a thermodynamic process generated by the quantum operation, $\Phi$.

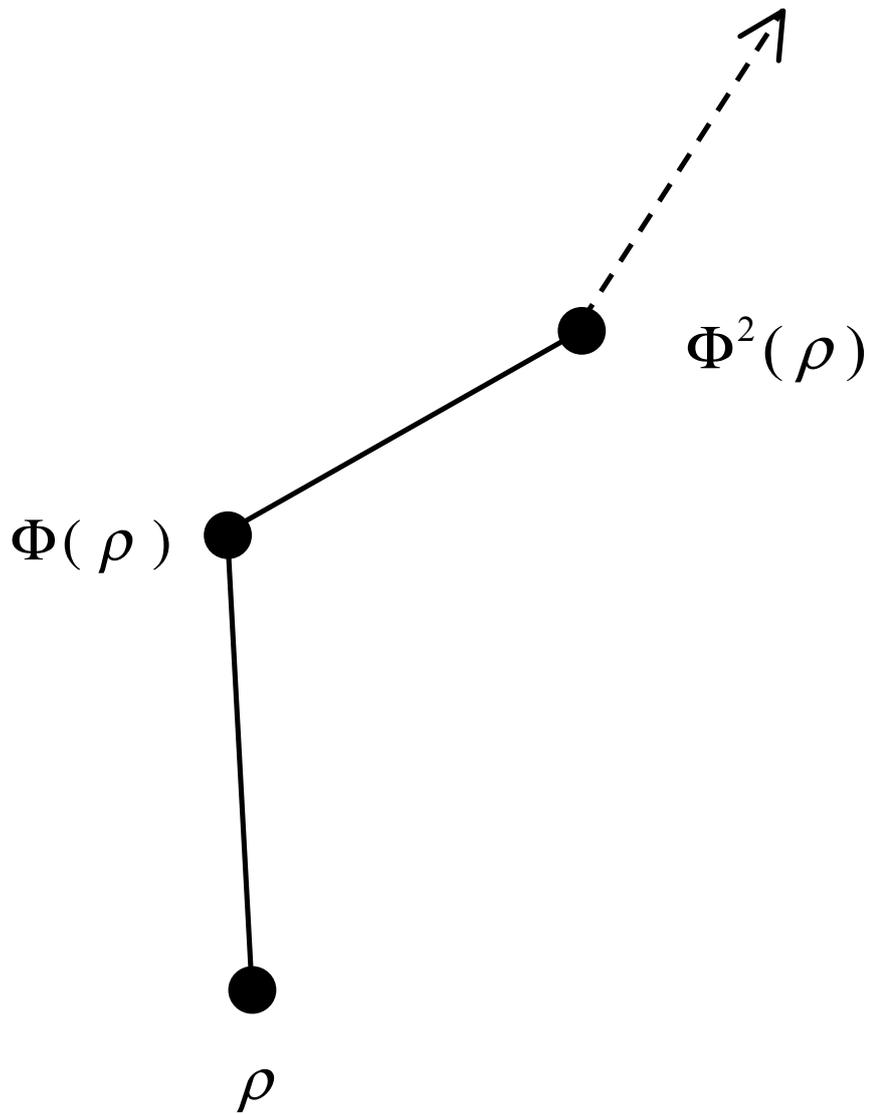

Fig. 1.